\documentclass[a4paper,11pt]{article}
\newcommand{\del}{\partial}

\title{Perturbative analysis on
      infrared aspects \\
       of noncommutative QED on $R^4$}
\author{M. Hayakawa
        \thanks{electric address: haya@post.kek.jp} \\
        {\it Theory Division, KEK,
             Tsukuba, Ibaraki 305-0801, Japan}   
       }
\date{December 18, 1999}
\begin{document}
\maketitle
\begin{abstract}
 Here we examine the noncommutative counterpart
of QED, which is called as noncommutative QED.
 The theory is obtained by examining the consistent
minimal coupling to noncommutative U(1) gauge field.
 The $*$-product admits the coupling of the matter
with only three varieties of charges, i.e., 0, +1 and $-$1.
 Ultraviolet divergence can be absorbed
into the rescaling of the fields and the parameters
at least at one loop level.
 To examine the infrared aspect of the theory
the anomalous magnetic dipole moment is calculated.
 The dependence on the direction of photon momentum
reflects the Lorentz symmetry violation of the system.
 The explicit calculation of the finite part
of the photon vacuum polarization
shows the infrared singularity
like ${\rm ln}({q\cdot C^TC\cdot q})$
($C^{\mu\nu}$ is a noncommutative parameter.)
which also exists in noncommutative Yang-Mills theory.
 It is associated with the ultraviolet behavior of the theory.
 We also consider the extension to chiral gauge theory
in the present context,
but the requirement of anomaly cancellation
allows only noncommutative QED.
\\
\\
{\it PACS\,}:\ 11.15.Bt, 11.10.Gh, 11.25.Db \\
{\it Keywords}:\ Noncommutative field theory;
                 Quantum electrodynamics; Perturbation theory;\\
\qquad \qquad \quad
                 Renormalization;
                 Anomalous magnetic dipole moment;
                 Chiral gauge theory
\end{abstract}
\section{Introduction}
\label{sec:intro}
\quad
  Noncommutative Yang-Mills (NCYM) theory
attracts one's interest after it emerged
in the BPS solution of Matrix theory \cite{Connes,Aoki_NCYM},
and is realized as the effective theory
around one of the perturbative vacua
of superstring theory with constant $B$ background
\cite{Douglas_B}.
 Prior to such recent development
noncommutative geometry and
the construction of field theory on it
has been developed \cite{Connes2}.
 Although the noncommutative geometry
appearing in the context of perturbative superstring theory
does not seem to have connection with quantum gravity,
the general noncommutative geometry
may accommodate the essential ingredients of quantum geometry
(which we do not know at present)
by its new machinery.
 Assuming the latter fact
we are naturally inclined to investigate
the quantum mechanical aspect of the theory
defined on the noncommutative geometry
in order to argue whether it really reflects
the microscopic structure of our world.
 But the first thing to do
is to analyze the simplest toy model and
know that the theory is well defined
even in the ultraviolet region.
 It is the most important subject
to find out how the infrared physics is affected,
and whether it provides more desirable foundation for
constructing the concrete model.
\\
\quad
 NCYM theory is widely discussed as it appears in string theory.
 The realization of such a theory
on the set of the ordinal functions
modifies the product of two functions
in terms of ``stared'' (mentioned hereafter as ``*''-)
product.
 Here we consider how the matter field
couples to Yang-Mills field in such a manner that
the theory respects this rule of product
as well as the gauge symmetry.
 The simplest candidate would be
a noncommutative counterpart of QED,
in which electron with a definite charge is present.
 As will be shown in Sec. \ref{sec:NC-QED}
inclusion of matters does not have so many varieties.
 In spite of the base manifold considered here
being topologically trivial,
charge is limited to three varieties; 0, $1$ and $-$1,
the precise meaning of which is defined
in Sec. \ref{sec:NC-QED}.
 This is stronger requirement
than the charge quantization on compact manifold
such as torus.
 We expect that there would be Hilbert space realization
for the theory with the fields of charge $\pm$ 1
on noncommutative geometry
and a mapping rule to
the function space on the usual coordinates
as was found in NCYM theory.
 Thus we call the fields with charge $+1$
as ``electron''
and the theory including such a object and ''photon''
as noncommutative QED (NC-QED) in this paper.
 Note that
the massive scalar fields receive quadratic divergence
which leads us to lose control of ultraviolet (UV) divergence
unless supersymmetry forbids its appearance.
 Contrarily
the usual fermion mass receives only logarithmic divergence
in four-dimensional QED or QCD.
 Nonlocal generalization of the interactions
is then expected not to drastically
modify the divergent structure
as experienced in NCYM system,
which also needs further clarification.
 Thus here the system involving fermions with charges
$\pm 1$ is examined primarily.
\\
\quad
 Although NCYM system is related to
ordinary Yang-Mills system by rather complex map
\cite{SW_NC},
there would be nontrivial quantum mechanical dynamics in NCYM
without maximal supersymmetry.
 One aspect of NCYM theory
has been argued to describe
the large N ordinary Yang-Mills theory with
a fixed t'Hooft coupling constant
in the high momentum region \cite{Bigatti,Ishibashi_NCYM}.
 It relies on
analysis of diagrams and
the pattern of momentum-dependent phase factors
appearing in it.
 More direct connection through the correlators of
the gauge invariant operators is welcome
in spite of curious nature of general Wilson loop operators
in noncommutative Yang-Mills theory
\cite{Ishibashi_NCYM,Maldacena_NCYM,Alishahiha_NCYM}.
 But it is beyond our present scope.
\\
\quad
 In order to access to the infrared side,
the first thing we can do is to investigate
the perturbative aspects.
 The perturbative analysis on UV structure of NCYM theory
has been done in Ref. \cite{Martin_NCYM,Krajewski}.
 The infrared side on NCYM as well as NC-QED
is our primary concern.
 Naive continuation of asymptotically free nature
\cite{Martin_NCYM} indicates
the existence of
such a dynamical scale as $\Lambda_{\rm QCD}$
at which the coupling constant diverges
while naive commutative limit reduces
to abelian gauge theory.
 The renormalization refers to the structures much higher than
noncommutative energy scale.
 Thus the commutative limit of the renormalized quantum theory
do not reduce it to its commutative counterpart
even in the low momentum region.
 We would like to begin in this paper
with seeking the dynamics
which could not be reached in local field theory,
by first examining a simple model, NC-QED.
\\
\quad
 The paper is organized as follows:
Sec. \ref{sec:NC-QED} is concerned with construction
of classical action involving the matter fields and showing
that the allowed choice is quite limited.
 In Sec. \ref{sec:infra} noncommutative QED theory
is quantized and the anomalous magnetic dipole moment
is calculated
to see whether the radiative effect from
noncommutative extension appears in the finite quantities.
 There the infrared behavior is further investigated
by observing the finite part of the vacuum polarization of photon.
 The potentially appearing $1/\tilde{q}^2$-singularity
cancels among the diagrams,
but the logarithmic infrared singularity is found,
which also exists in NCYM theory.
 Such a structure in the infrared side
is connected to the UV side.
 The extension to the chiral gauge theory is also
examined in Sec.\ref{sec:chiral},
but it is found
that there is {\it no} chiral gauge theory.
 Sec. \ref{sec:conc} is devoted to
the discussion and conclusion of the present paper.
\section{Noncommutative QED}
\label{sec:NC-QED}
\quad
  Pure noncommutative U(1) Yang-Mills action
\begin{equation}
 S_{YM} = \int d^d x\,
          \left( -\frac{1}{4 g^2} \right)
           F_{\mu\nu} *F^{\mu\nu}
          \, ,
 \label{eq:NYM_action}
\end{equation}
(Space-time dimension $d$ is set equal to four in final.)
is nothing but the one obtained from
the ordinary SU(N) Yang-Mills action
by replacing the matrix multiplication
to the ``star'' (hereafter referred as $*$-)product
\begin{equation}
 A * B(x) \equiv
 \left.
  e^{\frac{1}{2i} C^{\mu\nu}
     \del^{(\xi)}_\mu \del^{(\eta)}_\nu}\,
  A(x+\xi)\, B(x+\eta)
 \right|_{\xi,\eta \rightarrow 0} \, ,
\end{equation}
with an antisymmetric matrix $C^{\mu\nu}$
which characterizes noncommutativity of space-time
by modifying the algebra of functions.
 Even in U(1) case $A_\mu$ couples to itself
since the field strength $F_{\mu\nu}$ of $A_\mu$
has the nonlinear term
\begin{equation}
 F_{\mu\nu} = \del_\mu A_\nu - \del_\nu A_\mu
              -i [A_\mu,\, A_\nu]_{\rm M}\, ,
\end{equation}
where $[A,\, B]_{\rm M}$ denotes Moyal bracket:
\begin{equation}
 [A,\, B]_{\rm M} = A * B - B * A\, .
\end{equation}
 The $*$-product obeys the associative law
which is also satisfied by the matrix algebra
so that the manipulation in this case
resembles
the one experienced in the calculus of matrix.
 Thus it is quite simple to see
the action (\ref{eq:NYM_action}) is invariant
under 
\begin{equation}
 A_\mu(x) \rightarrow A^\prime(x)
  = U(x) *A_\mu(x) *U^{-1}(x) + i U(x) *\del_\mu U^{-1}(x)\, ,
  \label{eq:gauge_A}
\end{equation}
where $U(x) = (e^{i\theta(x)})_*$ is defined
by an infinite series expansion of multiple $*$-product of
scalar function $\theta(x)$.
 $U^{-1}(x) = (e^{-i\theta(x)})_*$
is similarly defined and plays the role
of the inverse of $U(x)$.
 Here and hereafter we assume
that the fields decrease so promptly at infinity
that the space-time integral of a Moyal bracket
(which corresponds to the trace of the commutator
in the ``matrix language'') vanishes.
\\
\quad
 The coupling of ``electron'' to gauge field
in noncommutative U(1) Yang-Mills theory
receives a severe restriction.
 The gauge transformation (\ref{eq:gauge_A}) for the gauge
field shows that
the simple candidate of the interaction
$\psi(x) * A_\mu(x)$ or $A_\mu(x) * \hat{\psi}$
implies that $\psi$ and $\hat{\psi}$ must transform as
\begin{equation}
 \psi(x) \rightarrow \psi^\prime(x) = U(x) *\psi(x)\, ,
 \quad
 \hat{\psi}(x) \rightarrow \hat{\psi}^\prime(x)
  = \hat{\psi}(x) *U(x)^{-1}\, ,
\end{equation}
respectively in order for each product
with gauge field to transform in the same
way as the original field.
 The covariant derivative
\begin{equation}
 D_\mu \psi = \del_\mu \psi - i A_\mu *\psi, \quad
 D_\mu \hat{\psi} = \del_\mu \hat{\psi}
  + i \hat{\psi} *A_\mu\, ,
\end{equation}
also transforms covariantly in the same way
as the original objects.
 Since the commutative limit leads to
the fields with $+1$ charge and $-1$ charge respectively
in ordinary QED,
so we call the field $\psi$ in the above
a field with $+1$ charge and referred hereafter as ``electron''
(opposite to the usual convention).
 Then the action
\begin{equation}
 S_{\rm matter} = \int d^d x
  \left(
   \bar{\psi} *\gamma^\mu iD_\mu \psi - m \bar{\psi} *\psi
  \right)
  \, ,
   \label{eq:matter_action}
\end{equation}
is invariant under local U(1) symmetry
since $\bar{\psi}$
\footnote{
 To compute the form factor
of the on-shell electron coupling to photon
we consider a theory in Minkowski space.
 Thus $\bar{\psi} = \psi^\dagger \gamma^0$.
}
behaves in the same manner as $\hat{\psi}$.
 The field with charge $+1$($-1$) in noncommutative case
would correspond to (anti-)fundamental representation
in ordinary nonabelian gauge theory.
 It is also reminiscent of such features that
noncommutative gauge theory carries
the internal degrees of freedom by imbedding them
into the space-time geometry itself.
 This is the reverse process of the reduction
of the space-time degrees of freedom
into the internal ones in the large N gauge theory
\cite{Kawai_RM}.
 When we pursue this correspondence further,
we are inclined to guess that
the higher-rank representation of SU(N) gauge theory
may convert into some matter fields
in noncommutative gauge theory.
 It would be the counterpart of the fields
of integral multiple of unit charge
from the view point of
noncommutative generalization of U(1) gauge theory.
 Actually the adjoint representation
corresponds to a field $\chi(x)$ with zero charge in total
but transforming in the by-product form
\begin{equation}
 \chi(x) \rightarrow \chi^\prime(x) =
      U(x) *\chi(x) *U^{-1}(x)\, .
\end{equation}
 Its covariant derivative is given by Moyal bracket.
 However we cannot find the counterpart
of the second-rank antisymmetric representation, etc,
of SU(N) gauge theory.
 The $*$-product admits only the fields with
charge $0$, $+1$ or $-1$.
 As a by-product the vacuum expectation value
of Wilson loop operator for a rectangular loop becomes
associated with the ground state energy
acting between two sources of charges, +1 and $-$1,
as usual.
\section{Perturbative Aspects in Infrared Side}
\label{sec:infra}
\quad
 We are interested in the quantum mechanical aspect of
the theory defined by the sum of (\ref{eq:NYM_action})
and (\ref{eq:matter_action})
\begin{equation}
 S_{\rm NC-QED}
  = \int d^d x\,
    \left(
     -\frac{1}{4 g^2} F_{\mu\nu} *F^{\mu\nu}
     +
     \bar{\psi} *\gamma^\mu iD_\mu \psi - m \bar{\psi} *\psi
   \right)
  \, .
 \label{eq:NC-QED}
\end{equation}
 Here we consider the theory on Minkowski space
with nonzero $C^{23}$ but vanishing $C^{01}$
in the canonical basis of antisymmetric matrix $C^{\mu\nu}$
for the later purpose of calculating
the anomalous magnetic dipole moment.
\\
\quad
 Perturbation theory begins with rescaling
$A_\mu \rightarrow g A_\mu$ and gauge fixing.
 BRST quantization as in ordinary QCD theory
leads the gauge fixing and Faddeev-Popov (FP) terms
\begin{equation}
 S_{\rm GF} =
  \int d^d x
  \left(
   -\frac{1}{2\alpha} \del_\mu A^\mu *\del_\nu A^\nu
   +
   \frac{1}{2}
   \left(
    i\bar{c} *\del^\mu D_\mu c -
    i\del^\mu D_\mu c * \bar{c}
   \right)
  \right)\, .
   \label{eq:NC-QED-GF}
\end{equation}
 Quantization is defined by perturbative expansion
due to Feynman rule as was done for NCYM theory
in Ref. \cite{Martin_NCYM},
but now derive from the actions (\ref{eq:NC-QED})
and (\ref{eq:NC-QED-GF})
\\
\quad
 The extra ultraviolet divergent contributions arise
in addition to those already appearing in NCYM theory.
 As we concentrate on reporting on the infrared phenomena
in this short article,
we postpone to describe the detail
about ultraviolet divergence at one-loop level
in the future extended volume \cite{Hayakawa},
and state the results only in brief:
 All the one loop ultraviolet divergence
can be subtracted by the local counterterms
with maintaining the equalities among
various $Z$ factors
(wave function renormalization constant, etc.)
required from gauge invariance.
 The $\beta$ function becomes for $N_F$ number of copies
of electron fields
\begin{equation}
 \beta(g) = \frac{1}{g} Q \frac{dg}{dQ}
  = -\left(
       \frac{22}{3} - \frac{4}{3} N_F
     \right)\, \frac{g^2}{16\pi^2}\, .
   \label{eq:beta}
\end{equation}
 A contribution $\frac{22}{3}$ is due to the structure similar
to nonabelian dynamics,
SU(2) Yang-Mills theory \cite{Martin_NCYM}.
 However the matter contribution is that found
in ordinary QED theory
with unit charge, {\it not} that of the quarks belonging to
the fundamental representation of SU(2) gauge theory
(in which $\frac{2}{3}$ instead of $\frac{4}{3}$ per flavor
found in (\ref{eq:beta})).
\\
\quad
 In this analysis
there is an important feature that should be kept in mind
for the analysis of infrared aspect of the theory.
 Only {\it planar} diagrams can have overall divergence.
 Noncommutativity of the theory manifests itself
in the form of a momentum-dependent phase
associated with the vertex in Feynman rule.
 ``Planar'' diagram is
a portion of the contributions
which has a definite phase factor,
but it only depends on the external momenta,
not on any loop momenta.
 Once loop momentum enters in the phase factor,
the suppression factor which depend
on the external momentum through $C^{\mu\nu}$
is always induced.
 This is the same feature already shared
by NCYM theory \cite{Filk,Bigatti,Ishibashi_NCYM}.
 The explicit two-loop computation similar to
Ref. \cite{Arefeva} in $\phi^4$ theory is welcome
to reveal  further detail structure
of the present theory.
\\
\quad
 To observe an infrared aspects of the theory,
the leading correction to magnetic dipole coupling
is estimated.
 The extraction of dipole coupling in the
$\psi\bar{\psi}A_\mu$ vertex function yields
\begin{eqnarray}
 i\left. \Gamma^\mu(p_I, p_F, q)\right|_{\rm dipole}
 &=&
 i g^3
 \left[
  e^{\frac{i}{2} p_I \cdot C \cdot p_F}\,H(1,p,q)
 \right.
  \nonumber \\
 && \quad \quad
  +
 \left.
  e^{\frac{i}{2} p_I \cdot C \cdot p_F}\,H(0,p,q)
  -
  e^{-\frac{i}{2}p_I \cdot C \cdot p_F}\,H(1,p,q)
 \right]\, mi\sigma^{\mu\nu} q_\nu \, ,
 \label{eq:MDM_1}
\end{eqnarray}
where $q$ is the incoming photon momentum,
and $p$ is connected to the incoming electron momentum
$p_I$ and the outgoing electron momentum $p_F$ through
\begin{equation}
 p_I = p - \frac{q}{2}, \quad p_F = p + \frac{q}{2}\, .
\end{equation}
 The matrix $\sigma^{\mu\nu}$ is here
$\sigma^{\mu\nu}
 = \frac{i}{2} \left[ \gamma^\mu, \gamma^\nu \right]$.
 The function $H(\eta,p,q)$ appearing in (\ref{eq:MDM_1}) is
\begin{eqnarray}
 H(\eta,p,q) &=& \int_0^{\infty} id\alpha_0
  \int_0^{\infty} id\alpha_+ \int_0^{\infty} id\alpha_+
  \frac{1}{[4\pi\beta i]^2}
  \nonumber \\
  && \quad \quad
  \times 2
  \left(
   \frac{\alpha_+ + \alpha_-}{\beta}
   -
   \left(
    \frac{\alpha_+ + \alpha_-}{\beta}
   \right)^2
  \right)
   \nonumber \\
 && \quad \quad
   \times
   \exp \left[
         -i\frac{1}{\beta}
           \left\{
            (\alpha_+ + \alpha_-)^2 m^2
             + \alpha_+ \alpha_- (-q^2)
           \right.
        \right.
         \nonumber \\
 && \qquad \qquad \qquad \qquad
       \left.
           \left.
             - \eta (\alpha_+ + \alpha_-) (p \cdot \tilde{q})
             + \eta^2 \frac{\tilde{q}^2}{4}
           \right\}
        \right]\, ,
  \label{eq:ex_H}
\end{eqnarray}
where $\beta = \alpha_0 + \alpha_+ + \alpha_-$, and
$\tilde{q}^\mu = C^{\mu\nu} q_\nu$
has the dimension of length.
 $H(0,p,q)$ becomes $\frac{1}{16\pi^2m^2}$ for on-shell photon.
 $H(1,p,q)$ can be written in terms of
a modified Bessel function of the second kind $K_1(x)$
\cite{math_formula}
\begin{eqnarray}
 H(1,p,q)
  &=&
   \frac{1}{8\pi^2}
   \int_0^1 d\alpha_+ \int_0^{(1-\alpha_+)} d\alpha_-
   \frac{(\alpha_+ + \alpha_-) - (\alpha_+ + \alpha_-)^2}
        {(\alpha_+ + \alpha_-)^2 m^2
          + \alpha_+ \alpha_- (-q^2)}
    \nonumber \\
  && \qquad \qquad \qquad \qquad \qquad \quad
   \times
   e^{i(\alpha_+ + \alpha_-) (p\cdot \tilde{q})}\,
   x\,K_1(x)
   \, ,
\end{eqnarray}
where $x = (-\tilde{q}^2)
           \left\{
            (\alpha_+ + \alpha_-)^2 m^2
              + \alpha_+ \alpha_- (-q^2)
           \right\}$.
 Since $K_1(x) \sim \frac{1}{x}$ for $x\sim 0$
we can take $q^2$ and $\tilde{q}^2$
\footnote{
 Since we consider the situation that only $C^{23}$ is nonzero,
thus there is an on-shell photon with a finite spatial momentum.
}
to zero without confronting with any singularities.
 Thus, for $q^2=0$ and $\tilde{q}^2=0$,
$H(1,p,q)$ is equal to $H(0,p,q)$.
 Therefore the leading correction to the magnetic dipole moment
is the same for ordinary QED and NC-QED.
 But the non-zero photon momentum in the direction
transverse to $(2,3)$ plane is allowed.
 Eq. (\ref{eq:MDM_1}) shows
that the strength of magnetic dipole coupling
is affected for such a photon in general.
\\
\quad
 We examine the infrared behavior of
the renormalized vertex functions,
especially the vacuum polarization for photon.
 As the analysis is lacking even for NCYM theory
\footnote{
 See also Ref. \cite{Seiberg}.
},
the common contributions,
the FP-ghost loop, gauge boson loop are examined here.
 Taking Feynman gauge for simplicity,
they can be written in terms of Schwinger parameterization
\cite{Itzykson}
\begin{eqnarray}
 &&
 i\Pi^{\mu\nu}_{\rm ghost + 33}(q) =
 ig^2 \int_0^\infty id\alpha_+ \int_0^\infty id\alpha_-
  \frac{1}{(4\pi\beta i)^{d/2}}
  \exp\left[
        -i \frac{\alpha_+ \alpha_-}{\beta} (-q^2)
      \right]
   \nonumber \\
 && \qquad \quad \times
 \left[
   \left(
     1 - \exp\left[ -i \frac{1}{\beta} \frac{\tilde{q}^2}{4} \right]
   \right)
   \times
   \left\{
     g^{\mu\nu} \left(
                 (3d-4) i\frac{1}{\beta} +
                 \left(
                   5-2\frac{\alpha_+ \alpha_-}{\beta^2}
                 \right) q^2
                \right)
   \right.
 \right.
  \nonumber \\
 && \qquad \qquad \qquad \qquad \qquad \qquad \qquad \quad
 + \left.
    q^\mu q^\nu
     \left(
       (d-6) - 4(d-2) \frac{\alpha_+\alpha_-}{\beta^2}
     \right)
   \right\}
   \nonumber \\
 && \qquad \qquad \qquad
\left.
 + \exp\left[ -i \frac{1}{\beta} \frac{\tilde{q}^2}{4} \right]
   \times \frac{1}{\beta^2} \times
   \left\{
    - \frac{1}{2} g^{\mu\nu} \tilde{q}^2
    + (2-d) \tilde{q}^\mu \tilde{q}^\nu
   \right\}
 \right]\, ,
    \nonumber \\
 &&
 i\Pi^{\mu\nu}_{4}(q) =
 2(d-1) i g^2 g^{\mu\nu}
 \int_0^\infty id\alpha \frac{1}{[4\pi\alpha i]^{d/2}}
 \left(
  1 - \exp\left[ -i \frac{1}{\alpha}
                    \frac{\tilde{q}^2}{4}
          \right]
 \right)\, .
  \label{eq:VP}
\end{eqnarray}
where $\beta = \alpha_+ + \alpha_-$.
 The first quantity in eq. (\ref{eq:VP})
is the contributions from the ghost loop
and the gluon loop induced through the two trilinear
gauge couplings.
 The other quantity is due to
the quartic self-interaction of gauge boson.
 The evaluation is similar to that
found in Ref. \cite{Itzykson}.
 The exponential factor
$\exp[-i /(4\beta \tilde{q}^2)]$ acts as the cutoff for
the ultraviolet divergence.
 The latter quantity in eq. (\ref{eq:VP})
is calculated as:
\begin{equation}
 i\Pi^{\mu\nu}_4(q) = 
  i \frac{g^2}{16\pi^2} g^{\mu\nu} \frac{-24}{-\tilde{q}^2}\, ,
   \label{eq:VP_4}
\end{equation}
containing a hard singularity $1/\tilde{q}^2$.
 It would be cancelled by
the term from $\Pi^{\mu\nu}_{\rm ghost + 33}(q)$.
 To evaluate it we need to perform the integrals
\begin{equation}
 \int_0^\infty \frac{d\rho}{\rho^{n+1}}
  \exp\left( -\rho - \frac{1}{\rho} a^2 \right)
 = \left( -\frac{1}{2a} \frac{d}{da} \right)^n [2K_0(2a)]\, .
   \label{eq:int_Bessel}
\end{equation}
where $a^2$ is proportional to $\tilde{q}^2 q^2$ in
the present context.
 Using
the asymptotic behavior of $K_0(x)$ around $x\sim 0$
available in a mathematical literature \cite{math_formula}
we can derive the useful formula
\begin{eqnarray}
 \int_0^\infty \frac{d\rho}{\rho}
 \exp\left( -\rho - \frac{1}{\rho} a^2 \right)
 &=&
  -{\rm ln}(a^2)
     \left( 1 + a^2 + {\cal O}(a^4) \right)
  - 2 \gamma_E + (-2\gamma_E + 2) a^2
   \nonumber \\
 && \quad
  + {\cal O}(a^4)\, ,
  \nonumber \\
 \int_0^\infty \frac{d\rho}{\rho^2}
 \exp\left( -\rho - \frac{1}{\rho} a^2 \right)
 &=&
  {\rm ln}(a^2)
   \left(
    1 + \frac{1}{2} a^2 + {\cal O}(a^4)
   \right)
   \nonumber \\
 && \quad
  + \frac{1}{a^2}
  +
  \left(
   2 \gamma_E - 1
  \right)
  + \left( \gamma_E - \frac{5}{4} \right) a^2
  + {\cal O}(a^4)\, ,
   \nonumber \\
 \int_0^\infty \frac{d\rho}{\rho^3}
  \exp\left( -\rho - \frac{1}{\rho} a^2 \right)
 &=&
  -{\rm ln}(a^2)
   \left(
    \frac{1}{2} + \frac{1}{6} a^2 + {\cal O}(a^4)
   \right)
   + \frac{1}{a^4} - \frac{1}{a^2}
     \nonumber \\
 && \quad 
   + \left(
      -\gamma_E + \frac{3}{4}
     \right)
   + \left(
      -\frac{1}{3} \gamma_E + \frac{39}{36}
     \right) a^2 + {\cal O}(a^4)\, .
      \label{eq:int_formula}
\end{eqnarray}
where $\gamma_E$ is Euler constant.
 They allow us to compute the singularity
of $\Pi^{\mu\nu}_{\rm ghost + 33}(q)$ for small $\tilde{q}^2$.
 It has the singular terms in the infrared
\begin{equation}
 i \Pi^{\mu\nu}_{\rm ghost + 33}(q) \sim
 i \frac{g^2}{16\pi^2}
   \left\{
     g^{\mu\nu} \frac{24}{-\tilde{q}^2}
     +
     \frac{10}{3}
     \left(
      g^{\mu\nu} q^2 - q^\mu q^\nu
     \right) {\rm ln} (q^2 \tilde{q}^2)
     + 32 \frac{\tilde{q}^\mu \tilde{q}^\nu}{\tilde{q}^4}
     - \frac{4}{3} \frac{q^2}{\tilde{q}^2}
       \tilde{q}^\mu \tilde{q}^\nu
   \right\} \, .
    \label{eq:singular}
\end{equation}
 Therefore $1/\tilde{q}^2$-singularity
cancels in the sum of (\ref{eq:VP_4}) and (\ref{eq:singular})
\begin{equation}
 i \Pi^{\mu\nu}(q) \sim
 i \frac{g^2}{16\pi^2}
   \left\{
     \frac{10}{3}
     \left(
      g^{\mu\nu} q^2 - q^\mu q^\nu
     \right) {\rm ln} (q^2 \tilde{q}^2)
     + 32 \frac{\tilde{q}^\mu \tilde{q}^\nu}{\tilde{q}^4}
     - \frac{4}{3} \frac{q^2}{\tilde{q}^2}
       \tilde{q}^\mu \tilde{q}^\nu
   \right\} \, ,
    \label{eq:singular}
\end{equation}
which is consistent with Slavnov-Taylor identity.
 The nonplanar contribution would diverge
if the integral in eq. (\ref{eq:VP})
were evaluated with $\tilde{q}^2$ set equal to zero.
 The logarithmic infrared singularity ${\rm ln}(\tilde{q}^2)$
in (\ref{eq:singular}) reflects
the fact that UV divergence is at most logarithmic
\footnote{
 D. Bigatti and L. Susskind has argued this structure
of singularities \cite{Bigatti}.
}.
 In fact the coefficient $10/3$ of ${\rm ln}(\tilde{q}^2)$
is that appears in the wave function renormalization factor
of photon in NCYM theory
\begin{equation}
  \left. Z_A \right|_{\rm NCYM}
     = 1 +
       \frac{g^2}{16\pi^2} \frac{10}{3}
       \frac{1}{\varepsilon^\prime}\, ,
  \label{eq:Z_A}
\end{equation}
where $1 /\varepsilon^\prime = 1 /\varepsilon + \gamma_E
                               - {\rm ln}(4\pi)$
for the space-time dimension $d=4 - 2\varepsilon$.
 The combination $q^2 \tilde{q}^2$ for logarithmic
correction assures this.
%
\section{Chiral Gauge Theory}
\label{sec:chiral}
\quad
 Until now all the fermions are
assumed to be Dirac fermions.
 It is naturally tempted
to pursue the extension to chiral gauge theory.
 The classical analysis given in Sec. \ref{sec:NC-QED}
is irrelevant to the chiral property of fermion.
 Thus Weyl fermions can have the charge $+1$ or $-1$.
 The right-handed fermion with $+1$ charge
is easily seen to be replaced by its CP conjugate
(the left-handed) fermion also in the present context.
 The chiral gauge theory
simply implies that
the number of the left-handed fermions with $+1$ charge
is not equal to one with $-1$.
 The question is whether such a theory can
circumvents a triangle loop anomaly
to define a consistent quantum theory or not.
\\
\quad
 It is easy to see the triangle loop diagram
is planar.
 Once we remind the correspondence between the current theory
to ordinary nonabelian gauge system
in which the external momentum
plays the role of color in Yang-Mills theory,
the remained integral
is evaluated completely in the same manner
as encountered in ordinary nonabelian gauge theory
involving the fundamental and/or anti-fundamental Weyl fermions.
 From this observation,
the number of the left-handed fermions with
$-1$ charge has to match with the number of fermions
with $+1$ in the system
\footnote{
 We require that
the triangular loop contribution cancels with each other
for {\it all} momentum configuration.
 But it might be too strong requirement
for noncommutative SU(N) gauge theory
due to non-factorizability of color and phase factors,
as suggested by Y. Kitazawa.
}.
 Such a theory is vector-like, i.e.,
noncommutative QED considered until the previous sections.
\section{Conclusion}
\label{sec:conc}
\quad
 In this paper we attempt to find
the noncommutative analogue of QED
and argue the perturbative aspects
of its infrared dynamics.
 The anomalous magnetic dipole moment
does not change at leading order
for the photon moving in the direction
along which noncommutativity is irrelevant.
 However the form factor seems to indicate
the possible observation of Lorentz invariance SO(1,3)
by controlling the direction of photon
although the conventional environment of measurement averages
over the direction of photon.
 In order to discuss quantum aspects of the theory,
it would be the best
to calculate and investigate
the radiatively corrected cross sections
for the electron-positron annihilation,
M$\phi$ller scattering,
Compton, or photon-photon scattering processes
with explicitly specified helicities of the external states.
\\
\quad
 For the preparation of this analysis,
the finite part of vacuum polarization
of the photon is calculated to study more about
the low momentum behavior.
 The absence of $1/\tilde{q}^2$ is connected
with the absence of quadratic UV divergence.
 In more detail the magnitude of the infrared singularity
is the same as that of logarithmic UV divergence.
\\
\quad
 The requirement of anomalous diagrams being cancelled
in total is too strong for chiral gauge theory
to exist in the present context.
 It is an interesting and important subject
to pursue the possibility to relax this requirement.
\\
\\
{\bf \Large Note added}
\\
\\
\quad
 During preparation of the paper,
we find a preprint \cite{Seiberg} reported
a few days ago,
which discussed
the subject partly overlapping with the present paper.
 The result here coincides with that obtained there.
\\
\quad
\\
{\bf \Large Acknowledgements}
\quad
\\
\\
\quad
 The author thanks
especially S. Iso for discussion and suggestion
at frequent times and reading manuscript
several times.
 He also thanks L. Susskind
for pointing out mistakes in the previous version
of the manuscript,
and N. Ishibashi, Y. Kitazawa, K. Okuyama and F. Sugino
for learning about noncommutative theory,
and the colleagues at KEK for sharing common interests
in this theory and the various topics
at a weekly informal meeting.

%

\begin{thebibliography}{10}
%
\bibitem{MLi}
  M. Li, ``Strings from IIB matrices'',
   Nucl.Phys. B499 (1997) 149-158, hep-th/9612222.
%
\bibitem{Connes}
 A. Connes, M. R. Douglas and A. Schwarz,
 ``Noncommutative geometry and matrix theory:
   Compactification on tori'',
  JHEP {\bf 02}, 003 (1998),
  hep-th/9711162.
%
\bibitem{Aoki_NCYM}
 H. Aoki, N. Ishibashi, S. Iso, H. Kawai,
 Y. Kitazawa and T. Tada,
 ``Noncommutative Yang-Mills in IIB matrix model'',
 hep-th/9908141.
%
\bibitem{Douglas_B}
 M. R. Douglas and C. Hull,
 ``D-branes and the noncommutative torus'',
  JHEP {\bf 02}, 008 (1998),
  hep-th/9711165.
%
\bibitem{Connes2}
 A. Connes,
  {\it Noncommutative Geometry} (Academic Press, 1994).
%
\bibitem{SW_NC}
 N. Seiberg and E. Witten,
 ``String theory and noncommutative geometry'',
  JHEP {\bf 09}, 032 (1999),
  hep-th/9908142;
 K. Okuyama,
 ``A path integral representation of the map
   between commutative and  noncommutative gauge fields'',
   hep-th/9910138.
%
\bibitem{Bigatti}
 D. Bigatti and L. Susskind,
  ''Magnetic fields, branes and noncommutative geometry'',
   hep-th/9908056.
%
\bibitem{Ishibashi_NCYM}
 N. Ishibashi, S. Iso, H. Kawai and Y. Kitazawa,
 ``Wilson loops in noncommutative Yang-Mills'',
  hep-th/9910004.
%
\bibitem{Maldacena_NCYM}
 J. M. Maldacena and J. G. Russo,
 ``Large N limit of non-commutative gauge theories,''
 JHEP {\bf 09}, 025 (1999),
 hep-th/9908134.
%
\bibitem{Alishahiha_NCYM}
 M. Alishahiha, Y. Oz and M.M. Sheikh-Jabbari,
  ``Supergravity and large N noncommutative field theories,''
  JHEP {\bf 11}, 007 (1999),
  hep-th/9909215.
%
\bibitem{Martin_NCYM}
 C. P. Martin and D. Sanchez-Ruiz,
 ``The one-loop UV divergent structure of U(1) Yang-Mills theory
   on  noncommutative ${\rm R}^4$'',
   Phys.\ Rev.\ Lett.\ {\bf 83}, 476 (1999),
   hep-th/9903077.
%
\bibitem{Krajewski}
 M. M. Sheikh-Jabbari,
 ``One loop renormalizability of supersymmetric Yang-Mills
   theories on noncommutative two-torus'',
   JHEP 9906 (1999) 015, hep-th/9903107;
 T. Krajewski and R. Wulkenhaar,
 ``Perturbative quantum gauge fields
   on the noncommutative torus'', hep-th/9903187.
%
\bibitem{Kawai_RM}
 T. Eguchi and H. Kawai,
 ``Reduction of dynamical degrees of freedom
  in the large N gauge theory'',
  Phys.\ Rev.\ Lett.\ {\bf 48}, 1063 (1982).
%
 \bibitem{Hayakawa}
  M. Hayakawa, in preparation.
%
 \bibitem{Filk}
  T. Filk,
  ``Divergencies in a field theory on quantum space'',
  Phys.\ Lett.\ {\bf B376}, 53 (1996).
%
\bibitem{Arefeva}
 I. Ya. Aref'eva, D. M. Belov and A. S. Koshelev,
 ``Two-Loop Diagrams in Noncommutative $\phi^4_4$ theory'',
  hep-th/9912075.
%
 \bibitem{math_formula}
  {\it Handbook of Mathmatical Functions},
   edited by M. Abramowitz and I. A. Stegun (Dover, 1972).
%
 \bibitem{Seiberg}
  S. Minwalla, M. V. Raamsdonk and N. Seiberg,
   ``Noncommutative Perturbative Dynamics'',
     hep-th/9912072
%
 \bibitem{Itzykson}
  C. Itzykson and J-B. Zuber,
   {\it Quantum Field Theory} (McGraw-Hill, 1985).
\end{thebibliography}
\end{document}